\documentstyle[12pt]{article}
\begin{document}

\begin{center}
{\Large \bf Gauge Independent Reduction of a \\
Solvable Model with Gribov-Like Ambiguity}
\end{center}
\begin{center}
\vskip 0.5 in
{\bf R. Banerjee}\footnote {On leave of absence from S.N. Bose National 
Centre for Basic Sciences, Calcutta, India. E-mail: rabin@if.ufrj.br}\\
Instituto de Fisica \\
Universidade Federal do Rio de Janeiro \\
C.P. 68528 \\
21945-970, Rio de Janeiro (RJ) \\
Brasil \\
\end{center}
\begin{abstract}
We present a gauge independent Lagrangian 
method of abstracting the reduced space of a
solvable model with Gribov-like ambiguity, recently proposed by Friedberg,
Lee, Pang and Ren. The reduced space is found to agree with the explicit
solutions obtained by these authors. Complications related to gauge fixing
are analysed. The Gribov ambiguity manifests by a nonuniqueness in the 
canonical transformations mapping the hamiltonian in the afflicted gauge
with that obtained gauge independently. The operator ordering problem in
this gauge is investigated and a prescription is suggested so that the 
results coincide with the usual hamiltonian formalism using the Schr\"odinger
representation. Finally, a Dirac analysis of the model is elaborated. In
this treatment it is shown how the existence of a nontrivial canonical set
in the ambiguity-ridden gauge  yields the connection with the previous 
hamiltonian formalism.
\end{abstract}
\newpage
{\Large \bf I. Introduction}
\vskip 0.5 in
In a recent paper Friedberg, Lee, Pang and Ren \cite {FL} have proposed a
solvable model which exhibits a Gribov-like ambiguity \cite {G} that is
known to exist in the Coulomb gauge of quantum chromodynamics. From the
explicit solutions of the model, these authors have also shown that it is 
necessary to include all gauge-equivalent copies rather than adopting
Gribov's suggestion of accouting for only those configurations having
a positive Fadeev-Popov determinant. The explicit results given in \cite {FL}
were obtained in the Hamiltonian approach using the Schrodinger representation
and in the path integral approach using Feynman rules.
Subsequently, a BRST analysis of this
model was also carried out by Fujikawa \cite {F}. Since many facets of the
Gribov problem are clouded by the complications of a nonabelian gauge theory
like quantum chromodynamics, the explicit computations possible in the
solvable model \cite {FL} provide an insight that cannot be otherwise gained.

It is obvious that the model \cite {FL} under consideration is a gauge
theory otherwise the issue of Gribov ambiguity does not arise. The 
conventional approach to isolate the true (physical) degrees of freedom
from the unwanted (unphysical) ones, which is characteristic of a gauge
theory, is to fix a gauge. In the hamiltonian formulation one usually 
starts from the time axial gauge where 
the Cartesian basis in the Schr\"odinger representation is defined. 
Transition to other
gauges is achieved by coordinate transformations from the results in the 
time axial gauge \cite {L}. This was the approach adopted
in \cite {FL}. It was also shown that the mapping from the time axial gauge
to a particular gauge was not one-to-one which was a manifestation of
a Gribov-like ambiguity in that gauge.

There is, however, an alternative way \cite {GI}, \cite {K} of obtaining the 
reduced (physical) hamiltonian without fixing any gauge. In the hamiltonian
formulation this reduction is based on the Levi-Civita transformations 
\cite {LC}. The viability or admissibility of any gauge is then shown
by demanding canonical equivalence with the gauge independent result. 
Nevertheless, a constructive prescription for carrying out this gauge 
independent reduction is still lacking.
Recently we \cite {R} have developed a purely Lagrangian approach
of systematically reducing the degrees of freedom in a gauge independent
manner. The physical hamiltonian is then obtained directly from this reduced
Lagrangian. Gauge fixing can also be implemented and its results are
analysed by discussing the canonical equivalence with the gauge independent
computations. A positive feature of this approach is that the gauge independent
analysis clearly reveals the most {\it natural} choice for gauge.

In this paper, using our aforesaid methods \cite {R}, a gauge independent
Lagrangian reduction of the model \cite {FL} will be presented in section III.
The physical hamiltonian, which is obtained directly from this Lagrangian,
is expressed in terms of the independent canonical pairs.
Different gauge fixings will be considered and the hamiltonian in these 
gauges will be explicitly computed. It is shown that the canonical 
transformations mapping the hamiltonian in a particular gauge 
with the hamiltonian obtained gauge independently
does not possess a unique inverse. This is the manifestation of the Gribov-like
ambiguity. Incidentally this nonuniqueness can be exactly identified with the
nonuniqueness present in the coordinate transformations considered in the
usual hamiltonian approach \cite {FL}. Our analysis shows 
that all gauge copies must be treated equivalently
which is compatible with the proposal made in \cite {FL}.
Additionally, it is found that
the hamiltonian in the gauge which suffers from the Gribov problem is plagued
by ordering ambiguities. A definite ordering prescription is given such
that the gauge fixed hamiltonian reproduces the corresponding 
expression obtained in \cite {FL} using the Schr\"odinger representation.
As an alternative hamiltonian formalism we have analysed this model in section
IV employing Dirac's \cite {D} theory of constrained systems. The 
complications of the Gribov problem are now revealed by a nontrivial pair
of canonical variables in the relevant gauge. By comparing this nontrivial set
with the standard canonical pairs found in Gribov ambiguity free gauges,
it is possible to reconstruct exactly the coordinate transformations \cite {FL}
mapping the hamiltonians in the distinct gauges. This establishes the 
connection of our Dirac analysis with the hamiltonian formalism of \cite {FL}.
To illustrate our ideas in a simpler, yet highly relevant, setting the gauge
independent Lagrangian reduction of the Christ Lee model \cite {CL} has been
presented in section II. Indeed many of the physical concepts and algebraic
manipulations developed here will be useful for section III. Our concluding 
remarks are given in section V.

We now give a brief review of the method \cite {R} of gauge independently
reducing the degrees of freedom in a given Lagrangian. An application
to electrodynamics is also included which serves to illuminate several 
distinguishing features, particularly its close connection with the 
models \cite {CL}, \cite {FL} considered in subsequent sections.
From the theory of differential equations unsolvable with
respect to the highest derivatives, it is possible to express the
lagrange equations for second order systems with variables $v$ by an
equivalent set of independent equations \cite {GI},
\begin{eqnarray}
\ddot p & =& \Theta (p, \dot p, q, \beta, \dot {\beta}, 
\ddot \beta) \label {1} \\
\dot q & =& \Phi (p, \dot p, q, \beta, \dot {\beta}) \label {2} \\
r & =& \Psi (p,  q, \beta) \label{3}
\end {eqnarray}
where $v = (p, q, r, \beta )$ and $\Theta, \Phi, \Psi $ are some
functions of the indicated arguments. Note that an overdot denotes a time
derivative. In a nonsingular theory $q, r,
\beta $ are absent so that there is an unconstrained dynamics with  
$ \ddot p = \Theta (p, \dot p)$. For singular theories (\ref {2}) and
(\ref {3}) represent the constraints. Now recall that the lagrange equations
were derived by a variational principle on the assumption that all
$v, \dot v$ were free. Since the constraints impose certain
restrictions on $v, \dot v$, it is essential that these keep the above set
of equations unmodified, or internal consistency is lost and the starting
Lagrangian is not valid.
Consequently time derivatives of the constraints must vanish by
virtue of this set of equations. This implies that the complete
constraint sector is given by (\ref {2}) and (\ref {3}).

    The idea is now to pass from the constrained $v = (p, q, r,
\beta)$ to the unconstrained $v = p$ by removing $q, r, \beta$. The
variable $r$ is trivially eliminated in favour of $p, q, \beta$ by
using (\ref {3}). In the physically interesting gauge systems the
constraints are implemented by a lagrange multiplier whose time derivative,
therefore, does not appear in the Lagrangian. This multiplier is
identified with $q$ which can thus be removed in favour of $p, \beta
$ by using (\ref {2}). The Lagrangian in the reduced sector is now a
function of $(v, \dot v; v= p, \beta) $. By evaluating the lagrange
equations in this sector it is possible to identify $\beta $ with the
variable that does not occur in these equations 
(see (\ref {1}) to (\ref {3})). With this
identification the variable $\beta $, which reflects the degeneracy
in the system, automatically drops out from the Lagrangian and its
final unconstrained form is obtained. The physical hamiltonian is now
found by performing the standard Legendre transformation.
Note that the usage of any gauge fixing has been completely avoided to 
obtain the reduced space. This analysis also indicates the most natural
choice of gauge as that which implies the vanishing of $\beta$. In that case
the reduced space obtained by gauge fixing would be trivially equivalent to
the gauge independent reduction. For an arbitrary gauge, however, it becomes
necessary to check the canonical equivalence between the gauge fixed and gauge
independent results, otherwise the gauge is not admissible.

An instructive illustration \cite {R} 
of this gauge independent Lagrangian reduction is provided by the
classic example of  spinor electrodynamics,
\begin{equation}
{\cal L} = -\frac{1}{4} F_{\mu\nu}F^{\mu\nu} + \bar \psi (i\partial
\!\!\!/\, -m
- eA \!\!\!\!/\,)\psi \label {6}
\end{equation}
The equations of motion are,
\begin{eqnarray}
(i\partial \!\!\!/\,-m -eA \!\!\!\!/\,)\psi &=& 0 \label {7} \\
\partial^\alpha F_{\alpha\mu}- ej_\mu &=& 0 \label {8}
\end{eqnarray}
where $j_\mu = \bar \psi \gamma_\mu \psi$ is the current. 
The $\mu =0$ component of (\ref {8}) is a
constraint. It can also be checked that this constraint is conserved in time
by virtue of the equations of motion. 
Furthermore there is a degeneracy in these equations
which becomes obvious from current conservation. The multiplier
$A_0$ (identified with $q$) can be eliminated in favour of the other
variables by solving the constraint. Using this, (\ref {6}) is
expressed in terms of the reduced set of variables. The Lagrange
equations in these variables are \cite {R},
\begin{equation}
\partial^j F_{ji} + \partial_0^2[(\delta_{ij} -
\frac{\partial_i\partial_j}{\partial^2})A_j] +
\frac{\partial_i}{\partial^2} \partial_0 j_0 - j_i = 0 \label{9}
\end{equation}
It is obvious that the variable $\beta$, manifesting the degeneracy,
is just the longitudinal $(L)$- component of $A_i$ which has dropped
out from (\ref {9}). Consequently by choosing the orthogonal polarisation,
\begin{equation}
A_i = A_i^T + A_i^L \label {x}
\end{equation}
the Lagrangian gets further reduced,
\begin{equation}
{\cal L} = \frac{1}{2} (\dot A_i^T)^2 - \frac{1}{4} F_{ij}^2 (A^T) +
\frac{1}{2} j_0\frac{1}{\partial^2}j_0 + j_i A_i^T +{\cal L}_M
\label{10} 
\end{equation}
where, expectedly, $A_i^L$ gets automatically removed and ${\cal
L}_M$ is the pure matter part. The unconstrained Lagrangian is expressed
in terms of the transverse $(T)$- component of $A_i$, which is the physical
(gauge invariant) variable.
The reduced Hamiltonian \cite {R} obtained from this Lagrangian
coincides with that found in the hamiltonian 
formalism \cite {GI} of abstracting the 
canonical set by a Levi-Civita \cite {LC}
transformation and then evaluating the total hamiltonian on the
constraint surface. It is now clear that the natural gauge choice would be the
Coulomb gauge $\partial_iA_i =0$, since this implies 
$A_i^L =0$. Indeed the Lagrangian (\ref {10}) obtained without gauge fixing
is exactly the Coulomb gauge-fixed
Lagrangian found in the literature \cite {GI}. The investigations
in the following sections provide further illustration and elaboration
of the gauge independent Lagrangian reduction.
\vskip 0.5 in
{\Large \bf II. The Christ Lee Model}
\vskip 0.5 in
A simple model put forward by Christ and Lee (CL)\cite {CL} has been useful
for testing different approaches \cite {K}, \cite {A}. Furthermore, since a 
straightforward generalisation of this model leads to that considered in
\cite {FL},
which will be discussed in the next section, it provides a convenient
starting point for the analysis. Indeed, as stated before, many physical
concepts and algebraic manipulations introduced here will also be exploited
in the next section. We begin with a gauge independent Lagrangian reduction
which is followed by a gauge fixed computation. The connection with the usual
hamiltonian formalism using the Schr\"odinger representation is established.

The CL model considers the motion of a point particle in two dimensional
space whose dynamics is governed by the Lagrangian,
\begin{equation}
L(x_i, \dot x_i, q) = \frac {1}{2} \dot x_i^2 - \epsilon_{ij}x_i\dot x_j q
+ \frac{1}{2}q^2 x_i^2 - V(\rho^2) 
\label{cl1}
\end{equation}
where $x_i = x_1, x_2$ are the rectilinear coordinates of the two dimensional
vector $\vec \rho$ so that $x_i^2= x_1^2 + x_2^2= \rho^2$, 
and $q$ is another coordinate whose time derivative is absent
in $L$ so that it plays a role analogous to the multiplier $A_0$ in QED. 
The antisymmetric tensor $\epsilon_{ij}$ is defined such that $\epsilon_{12}
= 1$. The
equations of motion obtained by varying $x_i$ and $q$ are given by,
\begin{eqnarray}
X_i &=& \ddot x_i + 2\epsilon_{ij}q\dot x_j + \epsilon_{ij}\dot q x_j
- q^2x_i +\frac {\partial V}{\partial x_i} = 0 \label {cl2} \\
Q & = & q\rho^2 - \epsilon_{ij}x_i\dot x_j = 0
\label {cl3}
\end{eqnarray}
There is a degeneracy or arbitrariness in these equations since,
\begin{equation}
\epsilon_{ij} x_i X_j + \dot Q = 0 \label {cl4}
\end{equation}
This is related to the invariance of the Lagrangian under the following
infinitesimal gauge transformations,
\begin{eqnarray}
x_i' & = & x_i - \theta (t)\epsilon_{ij}x_j \label {cl5} \\
q' & = & q + \dot \theta (t) \label {cl6}
\end{eqnarray}
This model, therefore, exhibits features similar to a gauge theory like
QED, except that the abelian group here comprises rotations instead of 
translations. Furthermore it is simple to identify the constraint, which
is the generator of gauge transformations, as being given by (\ref {cl3}). This
equation does not involve accelerations so that $q$ can be determined
from the initial Cauchy data. Moreover if (\ref {cl3}) is satisfied 
at one time,
it ramains valid at all times since $\dot Q = 0$ by virtue of (\ref {cl2})
and (\ref {cl4}). This brings out the exact analogy of $q$ with $A_0$, 
and that of (\ref {cl3}) with the Gauss constraint of QED. Following our
general strategy $q$ is now eliminated from (\ref {cl1}) by using (\ref {cl3})
to yield a reduced Lagrangian,
\begin{equation}
L(x_i, \dot x_i) = \frac {(x_i\dot x_i)^2}{2x_j^2} - V(\rho^2) \label {cl7}
\end{equation}
This is in exact analogy with the removal of $A_0$ in (\ref {6}) using 
the Gauss constraint.
It is now easy to see that this $L$ does not depend on $x_1$ and $x_2$
independently, but only on their combination $(x_1^2 + x_2^2)$. 
Introducing, therefore, the polar decomposition,
\begin{eqnarray}
x_1& =& \rho \cos\phi \nonumber\\
x_2& =& \rho \sin\phi \label {cl8}
\end{eqnarray}
it is clear that the redundant or cyclic variable that does not appear in
$L$ is $\phi$ whereas $\rho$ is the physical variable. In this variable we
obtain,
\begin{equation}
L(\rho, \dot \rho) = \frac{1}{2} {\dot \rho}^2 - V(\rho^2) \label {cl9}
\end{equation}
as the final {\it unconstrained} form of the Lagrangian. It just represents
the motion of the particle in one dimension subjected to the potential 
$V(\rho^2)$. The reduced hamiltonian is given by,
\begin{eqnarray}
H(\rho, \pi_\rho)& =& \pi_\rho\dot\rho - L(\rho, \dot\rho) \nonumber\\
& = &\frac{1}{2}{\pi_\rho}^2 + V(\rho^2) \label {cl10}
\end{eqnarray}
where $\pi_\rho = \dot\rho$ is the momentum conjugate to $\rho$.

This completes the gauge independent reduction of the CL model within a 
purely Lagrangian approach. We have found the physical hamiltonian
expressed in terms of the canonical variables. At this juncture it is
worthwhile to make certain comments. The process of eliminating the unphysical
or redundant variables is similar in both QED and the CL model. The Lagrange
multiplier $A_0$ or $q$ is first removed by exploiting the constraint. Then
the cyclic coordinate $A_i^L$ or $\phi$ is identified which automatically 
drops out from the reduced Lagrangian without any gauge fixing. To complete the
analogy, $\rho$ and $\phi$ in the CL model play the roles of $A_i^T$ and
$A_i^L$, respectively, in QED. Nevertheless, inspite of these similarities,
there is a basic difference in the manner in which the cyclic coordinate is
revealed. In QED an orthogonal polarisation (\ref {x}), 
which is effectively a way of
expressing $A_i$ in terms of shifted variables, is required whereas in the
CL model a curvilinear transformation (\ref {cl8}) has to be performed.
This is related to the fact that the abelian group in QED is translational
while it is rotational in the other case.

Let us next consider the issue of gauge fixing. Making the standard choice
of gauge \cite {K},\cite {A},
\begin{equation}
x_2 = \lambda x_1 \label {cl11}
\end{equation}
where $\lambda$ is a real parameter, we find the corresponding solution
for the multiplier $q$ from (\ref {cl3}),
\begin{equation}
q = 0 \label {cl12}
\end {equation}
since $x_1 = x_2 = 0$ is a singular point. The reduced Lagrangian following
from the simultaneous imposition of (\ref {cl11}) and (\ref {cl12}) in 
(\ref {cl1}) is given by,
\begin{equation}
L = \frac{1}{2} (1+\lambda^2)\dot x_1^2 - V((1+\lambda^2)x_1^2) \label {cl13}
\end{equation}
The canonical momenta is,
\begin{equation}
\pi_1 = \frac{\partial L}{\partial \dot x_1} = 
(1+\lambda^2)\dot x_1 \label {cl14}
\end{equation}
and the hamiltonian is found by a simple Legendre transform,
\begin{equation}
H = \pi_1 \dot x_1 - L = \frac{1}{2} \frac{\pi_1^2}{1+\lambda^2}
+ V((1+\lambda^2)x_1^2) \label {cl15}
\end{equation}

If we now make the following canonical transformation,
\begin{eqnarray}
x_1 & =& (\sqrt {1+\lambda^2})^{-1}\rho \nonumber\\
\pi_1 &=& (\sqrt {1+\lambda^2}) \pi_{\rho} \label {cl16}
\end{eqnarray}
where $(\rho, \pi_{\rho})$ constitutes the new canonical pair, it is seen
that the hamiltonian (\ref {cl15}) reduces to the expression found in
(\ref {cl10}). Furthermore the canonical transformation (\ref {cl16}) is
nonsingular and the inverse transformation can be trivially read-off. Thus
the dynamics obtained from the gauge fixed approach is canonically equivalent
to that found in the gauge independent method. This implies that the above
choice (\ref {cl11}) of gauge is allowed.

It is now instructive to physically unravel the meaning of the gauge 
independent and gauge fixed analysis, including their canonical equivalence.
The first step in this direction is to introduce the momenta conjugate
to $x_i$ in (\ref {cl1}),
\begin{equation}
\pi_i = \frac{\partial L}{\partial\dot x_i} = \dot x_i + \epsilon_{ij}x_jq
\label {cl17}
\end{equation}
In terms of this momenta the constraint (\ref {cl3}) becomes,
\begin{equation}
Q = \epsilon_{ij}\pi_i x_j = 0 \label {cl18}
\end{equation}
This implies the vanishing of the angular momentum so that the motion of
the particle is confined along a straight line. As far as 
the physics is concerned
the slope of this line is inconsequential. The gauge independent analysis
elegantly reproduces this dynamics by systematically removing the angular
variable $\phi$, thereby yielding the reduced hamiltonian (\ref {cl10}).
Now choosing the gauge (\ref {cl11}) merely fixes the slope of this line
as $\tan\phi = \lambda = \frac{x_2}{x_1}$. Consequently the dynamics in this
gauge is canonically equivalent to that obtained gauge independently. Indeed
the magnitude of the vector $\vec \rho$ (\ref {cl8}), 
expressed in terms of the
rectilinear coordinates $x_1$ and $x_2$ in the above gauge, is given by,
\begin{equation}
\rho^2 = x_1^2 + x_2^2 = (1 + \lambda^2)x_1^2 \label {cl19}
\end{equation}
which is just the inverse canonical transformation defined in (\ref {cl16}).
This completes our analysis of the CL model. Note that in the entire discussion
the choice of a specific Cartesian basis has been avoided since it was not 
necessary to introduce the Schrodinger representation. Nevertheless, as shown
below, it is simple to illustrate the connection with the 
conventional hamiltonian approach \cite {L} 
of choosing a Cartesian basis in the 
`time-axial' gauge and subsequently passing to other gauges by appropriate
gauge transformations.

The time-axial gauge in the CL model corresponds to taking $q=0$. In this gauge
the hamiltonian reduces to,
\begin{equation}
H = \frac{1}{2}\pi_i^2 + V(\rho^2) \label {cl20}
\end{equation}
where, using (\ref {cl17}), $\pi_i =\dot x_i$ with $x_1, x_2$ 
being the Cartesian
coordinates. As usual, the Gauss constraint (\ref {cl18}) does not emerge
from the analysis but has to be imposed by hand on the physical states.
This constraint can be solved by using the curvilinear transformation
(\ref {cl8}), with the consequence that the physical states become independent
of $\phi$. Correspondingly, the hamiltonian (\ref {cl20}) in these variables
has the form \cite {L},
\begin{equation}
H= -\frac{1}{2\rho}\frac{\partial}{\partial\rho}\rho \frac{\partial}{\partial
\rho} + V(\rho^2) \label {cl21}
\end{equation}
This result is the analogue of (\ref {cl10}) 
with the identification $\pi_\rho^2
\rightarrow -\frac{1}{\rho}\frac{\partial}{\partial\rho}\rho 
\frac{\partial}{\partial\rho}$.

Let us next consider the transition from the time-axial $(q=0)$ gauge
with cartesian coordinates $x_1, x_2$ to the gauge (\ref {cl11}) with
coordinates $X_1, X_2 (q\neq 0)$ brought about by (\ref {cl5}) and 
(\ref {cl6}),
\begin{eqnarray}
X_1&=&x_1 \cos\theta - x_2 \sin\theta \label {cl22} \\
\lambda X_1&=& x_1 \sin\theta + x_2 \cos\theta \label {cl23} \\
q& =& \dot \theta \label {cl24}
\end{eqnarray}
Equations (\ref {cl22}), (\ref {cl23}) define the coordinate trnsformations
from $(x_1, x_2)$ to $(X_1, \theta)$. Using the Cartesian representation
in the former set and the above transformations it can be shown that the 
Gauss constraint (\ref {cl18}) in the latter $(X_1, \theta)$ basis is
proportional to 
$(\frac{\partial}{\partial\theta})$. Repeating the above steps for the
hamiltonian (\ref {cl20}) and dropping terms proportional to $(\frac{\partial}
{\partial\theta})$, which enforces the Gauss constraint, finally yields,
\begin{equation}
H= -\frac{1}{2}(1+\lambda^2)^{-1} \frac{1}{X_1}
\frac{\partial}{\partial X_1}X_1 \frac{\partial}{\partial
X_1}
+ V((1+\lambda^2)X_1^2) \label {cl25}
\end{equation}
which just reproduces (\ref {cl15}) where, as before, the momenta $\pi_1$
conjugate to $X_1$ is identified by $\pi_1^2 \rightarrow 
-\frac{1}{X_1}
\frac{\partial}{\partial X_1}X_1 \frac{\partial}{\partial
X_1}$. This shows the equivalence between the reduced space hamiltonian
obtained in the conventional formulation \cite {L} 
of doing a coordinate transformation
from the (Cartesian) $q=0$ gauge result and that found directly by the 
Lagrangian approach. It is, however, important to stress that contrary to the
hamiltonian approach using the coordinate transformations, the general method
of Lagrangian reduction presented here does not require any gauge fixing.
Furthermore the validity of any gauge fixed computation has to established
by revealing its canonical equivalence with the gauge independent result. For
the specific choice (\ref {cl11}) this was  explicitly elaborated.

\vskip 0.5 in
{\Large \bf III. Extended Christ Lee Model}
\vskip 0.5 in
An extended version of the CL model proposed recently by Friedberg, Lee,
Pang and Ren \cite {FL} 
provides an interesting example of a solvable model with a 
Gribov-like ambiguity. This model has been analysed by these authors
in details using the
path integral and hamiltonian formulations. In the latter approach a
"time-axial" gauge, in which the Cartesian basis was defined, was taken
as the starting point. Transition to other gauges was achieved by performing
curvilinear coordinate transformations from the time-axial gauge, exactly
as shown here in the previous section for the CL model. It was then 
demonstrated that the mapping from the time-axial gauge to the so-called
``$\lambda$ gauge" \cite {FL} 
was not one-to-one, thereby realising a situation analogous
to the well known Gribov \cite {G} 
ambiguity. Here we shall apply our ideas elaborated
in the preceding sections to provide a gauge independent way of abstracting
the reduced (physical) space of this model. The reduced space will then be
derived by choosing a gauge. As announced earlier, the Gribov ambiguity gets
exposed by a {\it non-uniqueness} in the canonical transformation mapping
the reduced hamiltonian in the "$\lambda$ gauge" with the gauge independent
result. Connection of our results with the usual 
canonical hamiltonian formalism based on coordinate transformations from the
time-axial gauge will be discussed.

The model under consideration is defined by a simple extension of (\ref {cl1})
brought about by the introduction of a third Cartesian coordinate $z$, such
that the new Lagrangian reads,
\begin{equation}
L(x_i, \dot x_i, \dot z,q) = \frac {1}{2} \dot x_i^2 - 
g\epsilon_{ij}x_i\dot x_j q
+ g^2\frac{1}{2}q^2 x_i^2 + \frac {1}{2} (\dot z - q)^2 - V(x_1^2+x_2^2) 
\label{e1}
\end{equation}
where a coupling parameter $g>0$ has also been explicitly inserted. The
equation of motion obtained by varying $x_i$ is identical to (\ref {cl2}),
except that $q$ gets replaced by $gq$. The other Lagrange's equations are
given by,
\begin{eqnarray}
Z &=& \ddot z - \dot q = 0 \label {e2} \\
\tilde Q &=& g^2 q x_i^2 - g\epsilon_{ij}x_i\dot x_j - \dot z + q = 0 
\label {e3}
\end{eqnarray}
where the second equation is the analogue of (\ref {cl3}). It is easy to see
that the Lagrange equations are degenerate because of the relation,
\begin{equation}
\epsilon_{ij} x_iX_j + Z + \dot {\tilde Q} = 0 \label {e4}
\end{equation}
which resembles the relation (\ref {cl4}) in the CL model. This degeneracy 
is connected to the invariance of (\ref {e1}) under the gauge transformations
defined by (\ref {cl5}), (\ref {cl6}) (with $\theta$ replaced by
$\frac {\theta}{g}$) and,
\begin{equation}
z' = z + \frac {\theta}{g} \label {e5}
\end{equation}

Exactly as happened for the CL model, the variable $q$ acts as a Lagrange
multiplier. It is determined from the initial Cauchy data by 
the constraint (\ref {e3}). Furthermore this constraint is fixed in time,
as expected, since $\dot {\tilde Q} = 0$ by virtue of the degeneracy 
condition (\ref {e4}) and the equations of motion (\ref {cl2}) and
(\ref {e2}). Thus, as before, we solve (\ref {e3}) to eliminate $q$ from
the Lagrangian (\ref {e1}). The solution is given by,
\begin{equation}
q = \frac {\dot z + g(\epsilon_{ij}x_i \dot x_j)}{1+g^2(x_1^2+x_2^2)} 
\label {e6}
\end{equation}
At this point we use results from the previous
section to simplify the algebra. Noticing that the original
rotational invariance in the $x_1- x_2$ plane of the CL model is still
preserved we use the polar decomposition (\ref {cl8}) in terms of which
the solution (\ref {e6})reduces to,
\begin{equation}
q = \frac {\dot z + g\rho^2\dot \phi}{1+g^2\rho^2} \label {e7}
\end{equation}
Using this result the reduced Lagrangian, expressed in terms of polar
variables, obtained from (\ref {e1}) is,
\begin{equation}
L = \frac{1}{2}[\dot \rho^2 + (1+g^2\rho^2)^{-1}\rho^2(\dot \phi - g\dot z)^2]
- V(\rho^2)
\label {e8}
\end{equation}
It is clear that $L$ does not depend on $\phi$ and $z$ independently, but
only on their combination $(\phi - gz)$. Consequently it is trivial to
identify the redundant or cyclic variable that does not appear in the 
Lagrangian. Introducing a new variable,
\begin{equation}
\Phi = \phi - gz \label {e9}
\end{equation}
the final reduced Lagrangian $L_r$ in terms of unconstrained
variables is given by,
\begin{equation}
L_r = \frac{1}{2}[\dot \rho^2 + (1+g^2\rho^2)^{-1}\rho^2(\dot \Phi)^2]
- V(\rho^2) \label {e10}
\end{equation}
The canonically conjugate momenta are given by,
\begin{eqnarray}
\pi_\rho &=& \dot \rho \label {e11} \\
\pi_{\Phi} &=& \frac {\rho^2}{1+g^2\rho^2}\dot \Phi \label {e12}
\end{eqnarray}
The reduced hamiltonian follows by taking a standard Legendre transform,
\begin{eqnarray}
H &=& \pi_{\rho}\dot \rho + \pi_\Phi \dot \Phi - L_r \nonumber \\
& = & \frac{1}{2}\pi_{\rho}^2 + \frac{1}{2}(g^2 + \frac{1}{\rho^2})\pi_\Phi^2
+ V(\rho^2) \label {e13}
\end{eqnarray}
This concludes the gauge independent way of obtaining the final reduced
hamiltonian expressed in terms of the independent canonical pairs $(\rho,
\pi_\rho)$ and $(\Phi, \pi_\Phi)$.

Coming next to the issue of gauge fixing, the gauge independent analysis
provides the most natural way of proceeding. Remembering that eq. (\ref {e9})
isolated the redundant variable it is clear that the gauge choice,
\begin{equation}
z = 0 \label {e14}
\end{equation}
would yield a reduced hamiltonian that is trivially equivalent to the
gauge independent result (\ref {e13}), with the correspondence
$\Phi \rightarrow \phi; \pi_\Phi \rightarrow \pi_\phi$. 
Exactly as happened with the Coulomb gauge
in electrodynamics, it transpires that fixing the $z=0$ gauge in this model
is equivalent to not fixing any gauge. \footnote {
In this sense (\ref {e13}) will alternatively be referred to as the
hamiltonian in the $z=0$ gauge.} It is noteworthy
that the authors of \cite {FL} have also found this choice (which they
have termed as space-axial gauge) to be a more convenient starting point
for their analysis than the conventional time-axial gauge $q=0$. The gauge
independent analysis illuminates the reason behind this convenience. 
Although not considered in \cite {FL}, there is another equally convenient
gauge. This is given by choosing,
\begin{equation}
\phi = 0 \label {e15}
\end{equation}
It is easy to see from (\ref {e9}) that this choice is on an identical footing
to the space axial gauge (\ref {e14}). The canonical variables in this
gauge are given by the canonical transformation,
\begin{eqnarray}
\Phi & =& -gz \label {e16} \\
\pi_\Phi & = & -\frac{1}{g}\pi_z \label {e17}
\end{eqnarray}
Furthermore it is simple to return to the original coordinates since
in this gauge $\rho^2 = x_1^2$ and $\pi_\rho =\pi_1$. 
The hamiltonian is now written down directly
from (\ref {e13}),
\begin{equation}
H =  \frac{1}{2}\pi_1^2 + \frac{1}{2}(1 + \frac{1}{g^2x_1^2})\pi_z^2
+ V(x_1^2) \label {e18}
\end{equation} 

It is instructive to deduce the above hamiltonian (\ref {e18}) starting from
a gauge fixed lagrangian by adopting the standard procedure. This will
also show how to deal with more complicated gauges. The gauge (\ref {e15})
is equivalent to taking,
\begin{equation}
x_2 = 0 \label {e19}
\end{equation}
In that case the solution for the multiplier (\ref {e6}) simplifies to,
\begin{equation}
q = \frac{\dot z}{1 + g^2 x_1^2} \label {e20}
\end{equation}
The reduced Lagrangian after eliminating $q$ from (\ref {e1}) reads,
\begin{equation}
L(x_1, \dot x_1) = \frac{1}{2} [\dot x_1^2 + \frac{g^2x_1^2\dot z^2}
{1+g^2x_1^2}] - V(x_1^2) \label {e21}
\end{equation}
This Lagrangian is now expressed in terms of the unconstrained variables.
The canonical momenta are,
\begin{eqnarray}
\pi_1 &=& \dot x_1 \nonumber\\
\pi_z & = & \frac{g^2x_1^2\dot z}{1+g^2x_1^2} \label {e22}
\end{eqnarray}
and the hamiltonian obtained by a Legendre transform is given by,
\begin{equation}
H =  \frac{1}{2}\pi_1^2 + \frac{1}{2}(1 + \frac{1}{g^2x_1^2})\pi_z^2
+ V(x_1^2) \label {e23}
\end{equation}
which is just identical to (\ref {e18}). A logical extension of the 
gauge (\ref {e19}), which follows from the rotational symmetry in the 
$x_1-x_2$ plane, would be to consider the choice defined by (\ref {cl11}).
Since this gauge has been analysed in the context of the CL model we
do not pursue it further. It suffices to comment that the reduced hamiltonian
is found to be canonically equivalent to the gauge independent result
(\ref {e13}), exactly as shown in the case of the CL model.

Let us now try to understand the implications of the above analysis
from physical arguments. The constraint (\ref {e6}) can be written as,
\begin{equation}
\pi_z + g\epsilon_{ij}x_i\pi_j = 0 \label {e24}
\end{equation}
where momenta conjugate to the original variables $x_i, z$ in the Lagrangian
(\ref {e1}) have been 
introduced. This implies that the $z$- component of the angular momentum
of the particle is simulated by the $z$-component of the momentum. It is 
characteristic of confining the motion to the $x_1-x_2$ plane. Furthermore,
it is clear that this is also equivalent to considering the motion in 
either the $x_1-z$ or $x_2-z$ planes since these just constitute a renaming
of the coordinate-axes. Finally, from the rotational symmetry of the
problem, it follows that an identical description can be obtained by
regarding the motion on any plane normal to the $x_1-x_2$ plane. All these
possibilities have been considered within the gauge independent and subsequent
gauge fixed computations. The gauge independent result 
could be trivially identified with the $z=0$ gauge, which corresponds to
the motion in the $x_1-x_2$ plane. Likewise the $\phi =0$ and (\ref {cl11})
cases were considered, which correspond to motions in the $x_1-z$ plane and
any plane normal to the $x_1-x_2$ plane, respectively. All this is highly
reminiscent of our discussion in the CL model. There is, however, an
important distinction. In the CL model the gauge (\ref {cl11}) fixed the
slope of the straight line, which was the trajectory determined from
either physical or gauge independent considerations. There was no other
possibility. Here, on the other hand, we have not yet exhausted all freedom.
It is possible to choose a gauge that forces the motion of the particle 
to lie on a plane slanted to the $x_1-x_2$ plane. Such a gauge is defined by,
\begin{equation}
z = \lambda x_1 \label {e25}
\end{equation}
for any positive $\lambda$. It is referred to as the $\lambda$ gauge 
\cite {FL}. The motion in this gauge cannot be
determined easily from physical reasoning. We shall therefore abstract the
reduced hamiltonian as done earlier for the $x_2 =0$ gauge, and then
analyse its connection with the gauge independent result (\ref {e13}).

The starting point is to take the Lagrangian (\ref {e8}) obtained after
the elimination of the multiplier $q$. In terms of the polar variables
the condition (\ref {e25}) reduces to,
\begin{equation}
z= \lambda \rho \cos\phi \label {e26}
\end{equation}
from which the time derivative of $z$ may be computed. Inserting this 
in (\ref {e8}) yields,
\begin{equation}
L(\rho, \dot \rho, \phi, \dot \phi) = \frac{1}{2}\dot \rho^2
+\frac{1}{2} \beta^{-1} [ \alpha^2\rho^2\dot \phi^2 - 2g\lambda\rho^2
\alpha \dot \rho \dot \phi \cos\phi + g^2\lambda^2\rho^2\dot \rho^2 \cos^2\phi]
\label {e27}
\end{equation}
where,
\begin{eqnarray}
\alpha & =& 1+g\lambda\rho \sin\phi \nonumber\\
\beta &=& 1+g^2\rho^2 \label {e28}
\end{eqnarray}
The above Lagrangian is now written in terms of the {\it unconstrained} 
variables.
The canonical momenta are given by,
\begin{eqnarray}
\pi_\rho &=& \frac{1+g^2\rho^2(1+\lambda^2\cos^2\phi)}{\beta}
\dot \rho - \frac{g\lambda\rho^2 \alpha \cos\phi}{\beta}\dot \phi 
\label {e29} \\
\pi_\phi &=& \frac{\alpha^2}{\beta}\rho^2\dot \phi - \frac{\alpha}{\beta}
g\lambda\rho^2\dot \rho \cos\phi \label {e30}
\end{eqnarray}
As happens for unconstrained variables, the velocities can be inverted,
\begin{eqnarray}
\dot \rho &=& \pi_\rho + \frac{g\lambda \cos \phi}{\alpha}\pi_\phi \label
{e31} \\
\dot \phi &=& \frac{g\lambda \cos\phi}{\alpha}\pi_\rho
+ \frac{g^2\lambda^2\rho^2\cos^2\phi +\beta}
{\rho^2\alpha^2}\pi_\phi \label {e32}
\end{eqnarray}
Using these expressions the reduced hamiltonian can be found from (\ref {e27}),
by taking a suitable Legendre transform,
\begin{equation}
H= \frac{1}{2}[\pi_\rho^2+ \frac{2g\lambda \cos\phi}{\alpha}\pi_\rho\pi_\phi
+ \frac{g^2\lambda^2\rho^2\cos^2\phi +\beta}{\rho^2\alpha^2}\pi_\phi^2]
+V(\rho^2)
\label {e33}
\end{equation}
This is the final hamiltonian in the $\lambda$ gauge written in terms of the
independent canonical pairs $(\rho, \pi_\rho)$ and $(\phi, \pi_\phi)$.
It is simple to verify that if $\lambda$ is set equal to zero, then the
hamiltonian (\ref {e13}) in the $z=0$ gauge is reproduced. This serves as a
consistency check. 

We now concentrate on the general form of the hamiltonian for arbitrary
$\lambda$. As has been stressed the viability of the gauge fixed result
must be verified by demonstrating its equivalence, modulo canonical
transformations, with the gauge independent analysis. Indeed, in this
particular case, it can be checked that the following canonical transformation,
\begin{eqnarray}
\pi_\rho &=& \pi_\rho +\frac{g\lambda \cos\phi}{\alpha}\pi_\phi \nonumber\\
\rho &=& \rho \nonumber\\
\pi_\Phi &=& \alpha^{-1}\pi_\phi \nonumber\\
\Phi &=& \phi -g\lambda \rho \cos\phi \label {e34}
\end{eqnarray}
maps the gauge independent result (\ref {e13}) 
with (\ref {e33}). In the above set the 
canonical pairs on the L.H.S. correspond to (\ref {e13}) while those on the
R.H.S. correspond to (\ref {e33}). It is now necessary to check that the above
set (\ref {e34}) is nonsingular by working out the corresponding inverse
canonical transformation. It is obvious that if the inverse mapping
$\phi \rightarrow \Phi$ can be found, the other transformations follow from
simple algebraic manipulations. Now in the weak coupling limit $g\rightarrow
0$, this inverse mapping is a trivial identity transformation. Consequently
the weak coupling limit of the $\lambda$ gauge poses
no problems and can be regarded on an identical footing as the other gauges.
For arbitrary coupling, however, the inverse mapping $\phi \rightarrow \Phi$
is tricky and has been analysed in great details in ref. \cite {FL}. 
The result is that if,
\begin{equation}
\rho < (\lambda g)^{-1} \label {e35}
\end{equation}
the mapping is unique. However in those cases when,
\begin{equation}
\rho > (\lambda g)^{-1} \label {e36}
\end{equation}
the mapping is no longer unique. This implies that although the mapping
$\Phi \rightarrow \phi$ is one-to-one, the inverse transformation is
one-to-many. Thus the inverse canonical transformation to (\ref {e34})
is nonunique. This is the manifestation of the Gribov ambiguity. 
Indeed it is precisely for this range (\ref {e36}) of the coupling parameter
that the occurrence of the Gribov-like ambiguity was noted in \cite {FL}.
It is, however, important to point out that the canonical equivalence of the 
reduced hamiltonian in this gauge with that obtained gauge independently
remains valid for all different mappings. In other words this equivalence
does not discriminate any specific mapping and treats them all on an
identical footing. Our results therefore provide an independent confirmation
of the proposal made in \cite {FL} to regard all gauge copies equivalently
and not to isolate, as suggested by Gribov \cite {G}, any special copies.

We now elucidate the connection of our results with the hamiltonian formalism
in the Schr\"odinger representation carried out in \cite {FL}, 
based on coordinate transformations from the time axial gauge in which the
Cartesian basis is defined. Using these transformations it was shown \cite {FL}
that the hamiltonian in the space axial gauge $z=0$ had the form,\footnote {
Henceforth, to avoid notational confusion, we shall refer to the angular 
variable in the $z=0$ gauge and the $\lambda$ gauge by $\Phi$ and $\phi$,
respectively. While the latter symbol has already been used in the discussion
from (\ref {e26}) to (\ref {e33}), the former is prompted by the comments made
in footnote 2.}
\begin{equation}
H= -\frac{1}{2\rho}\frac{\partial}{\partial\rho}\rho \frac{\partial}{\partial
\rho} -  \frac{1}{2}(g^2 + \frac{1}{\rho^2})\frac{\partial^2}{\partial
\Phi^2}
+ V(\rho^2) \label {e37}
\end{equation}
It is simple to observe the equivalence of this expression with our result 
(\ref {e13}) using standard identifications in the 
Schr\"odinger representation,
\begin{eqnarray}
\pi_\rho^2
&\rightarrow & -\frac{1}{\rho}\frac{\partial}{\partial\rho}\rho 
\frac{\partial}{\partial\rho} \nonumber\\
\pi_\Phi &\rightarrow & -i\frac{\partial}{\partial \Phi} \label {e38}
\end{eqnarray}
where, it may be racalled, the first of these mappings was also used in the 
analysis of the CL model.

A similar coordinate transformation from the time axial gauge also led
to the hamiltonian in the $\lambda$ gauge. The result was found to be
\cite {FL},
\begin{eqnarray}
H&=& -\frac{\alpha^{-1}}{2\rho}[\frac{\partial}{\partial \rho}\rho \alpha
\frac{\partial}{\partial \rho} +\frac{\partial}{\partial \rho}
\lambda \rho g \cos\phi \frac{\partial}{\partial \phi}+
\frac{\partial}{\partial \phi}
\lambda \rho g \cos\phi \frac{\partial}{\partial \rho} \nonumber\\
&+& \frac{\partial}{\partial \phi}\alpha^{-1}(g^2\rho +\rho^{-1}+
\lambda^2 \rho g^2 \cos^2\phi) \frac{\partial}{\partial \phi}] +V(\rho^2)
\label {e39}
\end{eqnarray}
where $\alpha$ has been defined in (\ref {e28}). The above result
can be put in the form,
\begin{eqnarray}
H&=&-\frac{1}{2\rho}
\left [ \left (\frac{\partial}{\partial \rho}+g\lambda \cos\phi \alpha^{-1}
\frac{\partial}{\partial \phi}\right )\rho
\left (\frac{\partial}{\partial \rho}+g\lambda \cos\phi \alpha^{-1}
\frac{\partial}{\partial \phi}\right )\right ]\nonumber\\ &-&
\frac{1}{2}\left (g^2 + \frac{1}{\rho^2}\right )\alpha^{-2}
\left (\frac{\partial^2}{\partial
\phi^2}- g\rho \lambda \cos\phi \alpha^{-1}
\frac{\partial}{\partial \phi}\right )+V(\rho^2) \label {e40}
\end{eqnarray}
It is now easy to realise that this expression follows directly from the
hamiltonian (\ref {e37}) in the $z=0$ gauge by making the change of variables,
\begin{eqnarray}
\frac{\partial}{\partial \Phi}& =& 
\alpha^{-1}\frac{\partial}{\partial \phi} \nonumber\\
\frac{\partial}{\partial \rho}&=&
\frac{\partial}{\partial \rho}+g\lambda \cos\phi \alpha^{-1}
\frac{\partial}{\partial \phi} \label {e41}
\end{eqnarray}
These relations are the analogues of our canonical transformations (\ref 
{e34}) which
map the hamiltonian computed in a gauge independent manner with the 
hamitonian in the $\lambda$ gauge $(z=\lambda x_1)$. Indeed, exploiting
this analogy, it is possible to define the operator ordering in (\ref {e33})
so that the quantum theory obtained from it is in 
one-to-one correspondence with the Schrodinger
representation result, choosing the Cartesian basis in the time axial
gauge. It is clear that the issue of operator ordering is relevant only
for the hamiltonian (\ref {e33}). The gauge independent result 
(\ref {e13}), for instance, is free
of any ordering ambiguities. Since the gauge fixed result could be obtained
from the gauge independent one using canonical transformations
(\ref {e34}), it is evident
that the origin of the ordering problem is contained in these transformations.
A simple inspection shows that this is ineed true. By comparing 
(\ref {e41}) with
(\ref {e34}), and recalling the identification (\ref {e38}), 
it is found that the operators
in (\ref {e34}) should be ordered such 
that the canonical momenta occur after the 
coordinate variables. In fact the transformations (\ref {e34}) 
are already written with
this prescription.
If we now reconstruct the hamiltonian in the $\lambda$ gauge from the gauge
independent expression (\ref {e13}) keeping the ordering in (\ref {e34}) 
intact, we obtain,
\begin{eqnarray}
H &=& \frac{1}{2}[\pi_\rho^2+ \frac{g\lambda \cos\phi}{\alpha}\pi_\phi\pi_\rho
+\pi_\rho\frac{g\lambda \cos\phi}{\alpha}\pi_\phi+ \nonumber\\
&+&\frac{g\lambda \cos\phi}{\alpha}\pi_\phi 
\frac{g\lambda \cos\phi}{\alpha}\pi_\phi
+ \frac{\beta}{\rho^2}\frac{1}{\alpha}\pi_\phi\frac{1}{\alpha}\pi_\phi]
+V(\rho^2)
\label {e42}
\end{eqnarray}
It is simple to see that the terms quadratic in either $\pi_\rho$ or
$\pi_\phi$ in the above hamiltonian are mapped to their corresponding
structures in (\ref {e40}) on using the Schr\"odinger representations
(\ref {e38}).
The cross terms involving the product of $\pi_\rho$ and $\pi_\phi$ may
now be identified by inspection. This shows that (\ref {e42}) is the quantum
hamiltonian corresponding to the classical expression (\ref {e33}) such that
compatibility with the Schr\"odinger representation form (\ref {e40}) is 
preserved.
\vskip 0.5 in
{\Large\bf IV. The Dirac Analysis}
\vskip 0.5 in
It is well known that, besides the hamiltonian formalism utilising the 
Schr\"odinger representation, there is an alternative hamiltonian
approach which is based on Dirac's \cite {D} analysis of constrained
systems. We discuss Friedberg et al's model in this context and show how
the complications in the $\lambda$- gauge arise. Since the time derivative
of $q$ does not appear in (\ref {e1}), the momentum conjugate to it is a 
constraint,
\begin{equation}
\pi_q = 0 \label {d1}
\end{equation}
In Dirac's nomenclature \cite {D}, this is a primary constraint. There is a
secondary constraint which is found by time-conserving (\ref {d1}) with the
hamiltonian, 
\begin{eqnarray}
H &=& \pi_i \dot x_i + \pi_z \dot z +\pi_q \dot q - L \nonumber\\
& = & \frac{1}{2}(\pi_i^2 +\pi_z^2) +q (\pi_z + g\epsilon_{ij}x_i\pi_j) 
+ V(\rho^2) \label {d2}
\end{eqnarray}
obtained by taking a Legendre transform of (\ref {e1}). This secondary 
constraint, enforced by the Lagrange multiplier $q$, has already 
appeared in (\ref {e24}).
There are no further constraints. These constraints are first class since
their Poisson brackets are involutive. Indeed, as recognised earlier,
the secondary constraint is just the generator of the gauge transformations 
(\ref {cl5}) and (\ref {e5}). 
Corresponding to the two constraints there are two gauge
fixing conditions. One of these conditions must involve $q$ so that it
has a nonvanishing bracket with (\ref {d1}). A simple choice is
provided by,
\begin{equation}
q = 0 \label {d3}
\end{equation}
We shall next modify the other gauge condition to first consider the $z=0$
gauge and then the $\lambda$ gauge $(z=\lambda x_1)$. 
In the first case the full
set of constraints is now given by (\ref {e24}), (\ref {d1}), 
(\ref {d3}) and $z=0$. Here the canonical 
pairs are easily identifiable. A straightforward computation of the 
Dirac brackets \footnote {These brackets are defined by the standard formula
(\ref {d12})} (denoted by a star) 
reveals that these are given by $(x_1, \pi_1)$ and 
$(x_2, \pi_2)$ because,
\begin{eqnarray}
\{x_i, \pi_j\}^*& =& \delta_{ij} \nonumber\\
\{x_i, x_j\}^*& =& \{\pi_i, \pi_j\}^* = 0 \label {d4}
\end{eqnarray}
The brackets involving the other variables are redundant since these drop
out from the physical hamiltonian $(H_p)$ 
which corresponds to the evaluation of (\ref {d2}) on the constraint shell
(\ref {e24}). It is given by,
\begin{equation}
H_p = \frac{1}{2}(\pi_1^2(1+g^2x_2^2)+ \pi_2^2(1+g^2x_1^2)
-2g^2x_1x_2\pi_1\pi_2)+V(\rho^2) \label {d5}
\end{equation}
It is straightforward to show the equivalence of this hamiltonian, modulo
canonical transformations, with the expression (\ref {e13}). 
These transformations are given by,
\begin{eqnarray}
x_1&=& \rho \cos \Phi \nonumber\\
x_2& =& \rho \sin \Phi \nonumber\\
\pi_1&=& \pi_\rho \cos \Phi -\frac{\pi_\Phi}{\rho}\sin \Phi \nonumber\\
\pi_2&=& \pi_\rho \sin \Phi +\frac{\pi_\Phi}{\rho}\cos \Phi \label {d6}
\end{eqnarray}
There is no ambiguity in obtaining the inverse canonical transformations
which are given by,
\begin{eqnarray}
\rho^2& =& x_1^2 +x_2^2 \nonumber\\
\Phi &=& \tan^{-1}\frac{x_2}{x_1} \nonumber\\
\pi_\rho &=& \frac{x_1\pi_1+x_2\pi_2}{\sqrt {x_1^2+x_2^2}} \nonumber\\
\pi_\Phi &=& \epsilon_{ij}x_i\pi_j \label {d7}
\end{eqnarray}
Interestingly, such transformations 
were mentioned earlier \cite {K} in the hamiltonian
reduction of the CL model that was based on the Levi-Civita approach.

We now consider the $\lambda$ gauge. The analysis in the original Cartesian
variables is quite cumbersome and not very illuminating. Taking a hint
from the previous analysis, however, the problem can be simplified. 
A change
of variables in the original Lagrangian (\ref {e1}) is made by using the
standard 
polar decomposition given by the the first pair of equations in (\ref {d6}).
It is now possible to compute the constraint (\ref {e24})
in these variables by
working anew the Dirac prescription. But a short cut can be taken by
realising that the corresponding momenta are just the last pair
of equations in (\ref {d7}). The constraint is therefore given by,
\begin{equation}
\pi_z+g\epsilon_{ij}x_i\pi_j =\pi_z + g\pi_\phi = 0 \label {d8}
\end{equation}
The complete set of constraints $\Omega_i =0$ in the $\lambda$ gauge is now
explicitly enumerated,
\begin{eqnarray}
\Omega_1 &=& \pi_q \nonumber\\
\Omega_2 &=& q \nonumber\\
\Omega_3 &=& z-\lambda\rho \cos\phi \nonumber\\
\Omega_4 &=& \pi_z +g\pi_\phi \label {d9}
\end{eqnarray}
Since the Dirac brackets are nontrivial it is useful to give some algebraic
details. The matrix of the Poisson brackets of $\Omega_i$ is given by,
\begin{eqnarray}
\Omega_{ij} &=& \{\Omega_i, \Omega_j \} \nonumber\\
& =& \left (\begin{array}{cccc}
0 & -1 & 0 & 0 \\
1 & 0 & 0 & 0 \\
0 & 0 & 0 & \alpha \\
0 & 0 & -\alpha & 0 \end{array} \right )\label {d10}
\end{eqnarray}
where $\alpha$ has been defined in (\ref {e28}). The inverse matrix is easily
computed,
\begin{eqnarray}
\Omega_{ij}^{-1} &=& (\{\Omega_i, \Omega_j \})^{-1} \nonumber\\
& =& \left (\begin{array}{cccc}
0 & 1 & 0 & 0 \\
-1 & 0 & 0 & 0 \\
0 & 0 & 0 &- \alpha^{-1} \\
0 & 0 & \alpha^{-1} & 0 \end{array} \right )\label {d11}
\end{eqnarray}
Dirac brackets among any two variables $A$ and $B$ can be calculated using
the formula \cite {D},
\begin{equation}
\{A, B \}^*= \{A, B \} -\{A, \Omega_i \}\Omega_{ij}^{-1}\{\Omega_j, B \}
\label {d12}
\end{equation}
The nontrivial Dirac brackets (i.e. those differing from the corresponding
Poisson brackets) are explicitly written,
\begin{eqnarray}
\{\phi, \pi_\phi \}^*& =&\alpha^{-1} \nonumber\\
\{\phi, \pi_\rho \}^*& =& (g\lambda \cos \phi )\alpha^{-1} \label {d13}
\end{eqnarray}
As a simple consistency check of this algebra, it is straightforward to
reproduce the results in the $z=0$ gauge by setting $\lambda =0$. In this
case $\alpha =1$ and the canonical pairs are $(\Phi, \pi_\Phi)$ and
$(\rho, \pi_\rho)$. This was realised earlier in the canonical transformations
(\ref {d6}) mapping (\ref {d5}) to (\ref {e13}).

To proceed for arbitrary $\lambda$, it is first essential to identify the
canonical set. This is done by exploiting the algebra,
\begin{eqnarray}
\{\phi -g\lambda\rho \cos\phi, \pi_\phi \}^*& =& 1 \nonumber\\
\{\phi -g\lambda\rho \cos\phi, \pi_\rho \}^*& =& 0 \label {d14}
\end{eqnarray}
deduced from the basic brackets (\ref {d13}). Consequently the independent 
canonical pairs in the $\lambda$ gauge are $(\rho, \pi_\rho)$ and
$(\phi -g\lambda\rho\cos\phi, \pi_\phi)$. With this knowledge it is feasible
to reconstruct the hamiltonian in the $\lambda$ gauge from the result in the
$z=0$ gauge. A comparison of the canonical sets in these two gauges
immediately leads to the following mappings in the Schrodinger representation,
\begin{eqnarray}
\Phi& =& \phi -g\lambda\rho\cos\phi \nonumber\\
\frac{\partial}{\partial \Phi}& =& 
\alpha^{-1}\frac{\partial}{\partial \phi} \nonumber\\
\rho &=&\rho \nonumber\\
\frac{\partial}{\partial \rho}&=&
\frac{\partial}{\partial \rho}+g\lambda \cos\phi \alpha^{-1}
\frac{\partial}{\partial \phi} \label {d15}
\end{eqnarray}
such that compatibility with (\ref {d14}) is preserved. The hamiltonian in 
the $z=0$ gauge was already shown to be canonically equivalent to (\ref {e13})
which, in the Schr\"odinger representation, is given by (\ref {e37}). With the
above mapping (\ref {d15}), the hamiltonian (\ref {e40}) in the $\lambda$
gauge is reproduced from (\ref {e37}). This becomes obvious if it is 
realised that (\ref {d15}) is identical to the mapping (\ref {e41}),
found on the basis of coordinate transformations, which was shown earlier
to connect the hamiltonian in the two gauges. Consequently the correspondence
between the two hamiltonian approaches, based on either Dirac's \cite {D}
analysis or using coordinate transformations in the Schr\"odinger
representation \cite {FL}, is established.

\vskip 0.5 in
{\Large \bf V. Conclusions}
\vskip 0.5 in
Using the gauge independent method, recently developed by us \cite {R},
of reducing the degrees of freedom in a gauge theory we have presented
a detailed analysis of a solvable model with Gribov-like ambiguity, proposed
by Friedberg, Lee, Pang and Ren \cite {FL} (called here as `Extended
Christ Lee Model'). This reduction was carried
out at the Lagrangian level by systematically eliminating the redundant or
unphysical degrees of freedom in a two-step process. 
First, the multiplier that enforces the Gauss constraint was eliminated by
solving this constraint. Next, a change of variables was effected which
eliminated the cyclic coordinate that was responsible for the degeneracy
in the equations of motion that is characteristic of any gauge theory.
This change of variables was dictated by the nature of the gauge symmetry.
For example in QED where this symmetry is translational, a simple
shift (\ref {x}) was required, whereas in either Friedberg et al's model
\cite {FL} or the Christ Lee (CL) model \cite {CL} where the symmetry is
rotational, a curvilinear transformation was necessary. In all instances
the cyclic coordinate was identified and naturally eliminated from the 
Lagrangian which was now expressed solely in terms of unconstrained
variables.The physical hamiltonian was computed directly
from this unconstrained Lagrangian.

The reduced space was also obtained by fixing a gauge whose effect was to
impose certain restrictions on the unphysical variables. The viability
of the gauge fixed computations was judged by checking the canonical 
equivalence with the results obtained gauge independently. In fact the
relevant canonical transformations were explicitly worked out for both the
CL model and Friedberg et al's model. Interestingly, in the latter case it
was found that the canonical transformations in the $\lambda$ gauge
$(z=\lambda x_1)$ did not possess a unique inverse, thereby yielding a 
Gribov-like phenomenon. Our analysis showed that all gauge copies must be
treated equivalently, which corroborated earlier findings \cite {FL},
\cite {F}. The connection of the canonical transformations discussed here
with the coordinate transformations in the usual hamiltonian formalism
employing the Schr\"odinger representation \cite {L} was clearly revealed
for either Gribov ambiguity free gauges or the $\lambda$ gauge. Using this
connection the operator ordering problem in the $\lambda$ gauge was resolved.

As an alternative hamiltonian formalism, Friedberg et al's model was also
investigated using Dirac's \cite {D} constrained analysis. The occurrence
of a nontrivial canonical set in the $\lambda$ gauge, as opposed to the set
in the $z=0$ gauge, led to a mapping relating the physical hamiltonians in
these gauges. In the Schr\"odinger representation this mapping was exactly
identical to that found by using Friedberg et al's analysis.

Coming back to the gauge independent analysis it provided a natural way of
understanding those reduction process in these models that were dictated from
purely physical arguments. Furthermore, a simple 
way of identifying a convenient gauge to be employed in gauge 
fixed computations
was an important consequence of the gauge independent analysis. Just as the
Coulomb gauge was shown to be the natural choice in QED, the $z=0$ gauge
was the corresponding choice in Friedberg et al's model. Significantly,
the authors of \cite {FL} have found this choice, instead of the conventional
time-axial gauge $(q=0)$, as a more convenient starting point for their 
hamiltonian formulation. In more complicated theories, therefore, the 
guideline provided by the gauge independent analysis concerning gauge fixing
could be valuable.
\vskip 0.4 in
{\Large \bf Acknowledgments}
\vskip 0.4 in
The author thanks the members of the IF, UFRJ, for their kind hospitality
and the CNPq (Brasilian National Research Council) for providing financial
support.

\end{document}